\documentclass[12pt]{article}

\usepackage{amsmath}
\usepackage{graphicx,psfrag,epsf, xcolor, tikz, bm}
\usetikzlibrary{backgrounds}
\usepackage{enumerate}
\usepackage{amsfonts}
\usepackage{amsmath,epsfig,amssymb,amsfonts,amsthm,epsfig,verbatim}
\usepackage{color,graphicx,lscape,longtable}
\usepackage{graphicx}
\usepackage{float}
\usepackage{subfigure}
\usepackage{indentfirst}
\usepackage{multirow}

\def\bSig\mathbf{\Sigma}

\usepackage{tikz}

\title{Sparse Estimation of Historical Functional Linear Models with a Nested Group Bridge Approach}
\date{}
\author
{
Xiaolei Xun\footnote{School of Data Science, Fudan University, Shanghai, China} \\
\textit{Email: xiaolei\_xun@fudan.edu.cn} \\
\\
Jiguo Cao \footnote{Department of Statistics and Actuarial Science, Simon Fraser University, Burnaby, BC, Canada} \\
\textit{Email: jiguo\_cao@sfu.ca}
}

\makeatletter
\renewcommand\@biblabel[1]{}
\renewenvironment{thebibliography}[1]
{\section*{\refname}%
	\@mkboth{\MakeUppercase\refname}{\MakeUppercase\refname}%
	\list{}%
	{\leftmargin0pt
		\@openbib@code
		\usecounter{enumiv}}%
	\sloppy
	\clubpenalty4000
	\@clubpenalty \clubpenalty
	\widowpenalty4000%
	\sfcode`\.\@m}
{\def\@noitemerr
	{\@latex@warning{Empty `thebibliography' environment}}%
	\endlist}
\makeatother

\begin{document}

\maketitle

\begin{abstract}
The conventional historical functional linear model relates the current value of the functional response at time $t$ to all past values of the functional covariate up to time $t$. Motivated by situations where it is more reasonable to assume that only recent, instead of all, past values of the functional covariate have an impact on the functional response, we investigate in this work the historical functional linear model with an unknown forward time lag into the history. Besides the common goal of estimating the bivariate regression coefficient function, we also aim to identify the historical time lag from the data, which is important in many applications. Tailored for this purpose, we propose an estimation procedure adopting the finite element method to conform naturally to the trapezoidal domain of the bivariate coefficient function. A nested group bridge penalty is developed to provide simultaneous estimation of the bivariate coefficient function and the historical lag. The method is demonstrated in a real data example investigating the effect of muscle activation recorded via the noninvasive electromyography (EMG) method on lip acceleration during speech production. The finite sample performance of our proposed method is examined via simulation studies in comparison with the conventional method. \\
\end{abstract}


\section{Introduction}
\label{sec:intro}

We are interested in functional linear regression models, where a functional predictor $x_i(t)$ is used to explain a functional response $y_i(t)$ over a time interval $[0, T]$, $i=1, \ldots, N$. 
There are a number of possible ways to build up the model. For instance, Ramsay and Dalzell (1991) introduced the functional linear model
\begin{equation}
	\label{mod:fun_on_fun}
	y_i(t) = \alpha(t) + \int_0^T \beta(s,t)  x_i(s)ds + \varepsilon_i(t), \quad t \in [0,T],
\end{equation}
where $\alpha(t)$ is the intercept function, $\varepsilon_i(t)$ is the residual and $\beta(s,t)$ is an unconstrained regression bivariate coefficient function of primary interest. The bivariate coefficient function $\beta(s,t)$ represents the effect of the functional predictor at time $s$ on the functional response at time $t$. According to model~(\ref{mod:fun_on_fun}), future facts about the covariate $x(s) \mbox{ with } s>t$ is used to explain $y(t)$, the response at current time. This is reasonable if the underlying process is periodic, but illogical otherwise.
If the prediction of the response $y(t)$ only depends on concurrently observed predictor $x(t)$, a restriction is then introduced on $\beta(s,t)$, leading to the functional concurrent model
$y_i(t) = \alpha(t) + x_i(t)\beta(t) + \varepsilon_i(t)$, for $t \in [0, T]$
where a univariate coefficient function $\beta(t)$ suffices to fully describe their relationship (Ramsay and Silverman, 2005). In other situations, the response at a particular time point may depend on the recent history of the predictor, then the functional concurrent model becomes too constrained. For example, the recovery of a patient may well depend on treatment received over the past a few days.

To properly model the relationship between the historical predictor and the current response, Malfait and Ramsay (2003) introduced the historical functional linear model
\begin{equation}
	\label{mod:historical}
	y_i(t) = \alpha(t) + 
	\int_{t-\delta}^t \beta(s,t) x_i(s) ds + \varepsilon_i(t),
	\quad t \in [0, T],
\end{equation}
where $\delta \in [0, T]$ represents a forward time lag back into the history such that $x_i(s)$ is likely to have an impact on $y_i(t)$ only over the time frame $[t-\delta, t]$ rather than over the complete time frame $[0, T]$. Notice that the degenerated case with $\delta=0$ reduces to a concurrent model. The time lag $\delta$ is typically unknown and of interest. The objective of our study is to estimate the time lag $\delta$ along with the coefficient function $\beta(s,t)$ from the data .

Concerning the classic function-on-function regression model (\ref{mod:fun_on_fun}), one can apply basis expansion to $x_i(t)$ and $y_i(t)$, and obtain a weighted least squares type estimator, possibly with a bivariate roughness penalty 
(Besse and Cardot, 1996; Ramsay and Silverman, 2005).
Alternatively, one can compute the functional principal component scores for $x_i(t)$ and $y_i(t)$, and base the estimation on the functional principal component scores (Chen and Wang, 2011; Ivanescu, et al., 2015; Yao, M\"{u}ller, and Wang, 2005). 

The functional concurrent model, as a type of varying coefficient model, has also been studied relatively well in the literature (see, Fan and Zhang, 2000; Hastie and Tibshirani, 1993;  Wu, Chiang and Hoover, 1998; Wu and Liang, 2004; Zhou, Huang, and Carroll, 2008).
For a comprehensive introduction to functional regression models, we refer to monographs by Ferraty and Vieu (2006), Hsing and Eubank (2015), Kokoszka and Reimherr (2017), Ramsay and Silverman (2005), as well as the review papers by Morris (2015) and Wang, Chiou and M\"{u}ller (2016) and references therein.

There also exist intermediate models between the functional linear model~(\ref{mod:fun_on_fun}) and the functional concurrent model which try to model the effect of past values of the predictor on current response. A class of time-varying functional regression models were discussed by M\"{u}ller and Zhang (2005), \c{S}ent\"{u}rk and M\"{u}ller (2008; 2010), etc.
Brockhaus et al. (2017) and Greven and Scheipl (2017) discussed a general framework of functional regression, where many aforementioned models are included as special cases, and proposed gradient-boosting-based estimation procedures.
Kim, \c{S}ent\"{u}rk and Li (2011) proposed an estimation procedure that was geared towards sparse longitudinal data with irregular observation times and small amount of measurements per subject.
%
Assuming the dependence of $y(t)$ on $x(s)$ for $s\in[t-\delta, t]$ does not change over time, Asencio, Hooker and Gao (2014) introduced the functional convolution model which is a functional extension of distributed lag models in time series, 
and proposed a penalized ordinary least squares estimator for the regression coefficient given the historical time lag $\delta$.

The historical functional linear model~(\ref{mod:historical}) is identifiable but not estimable, with effectively an infinite number of covariates, therefore regularization or roughness penalty on $\beta(s,t)$ is necessary in the estimation process (Ramsay and Silverman, 2005). 
Malfait and Ramsay (2003) regularized the fit by approximating $\beta(s,t)$ using an expansion with a finite number of basis functions. Such approach bears the well known limitation that a small number of basis functions would decrease the goodness of fit while a large number of basis functions would lead to unstable estimation. 
Harezlak et al. (2007) improved the fit of model~(\ref{mod:historical})  by imposing a discrete roughness penalty forcing neighboring coefficients to be similar, which is an extension of the P-spline by Eilers and Marx (1996).
%
%
Both work focus on the estimation of coefficient function $\beta(s,t)$ for a given historical lag $\delta$, while the lag $\delta$ is decided in a separate manner. Malfait and Ramsay (2003) considered deciding $\delta$ as a model selection problem. Briefly put, they divided the whole time course $[0,T]$ into $M$ sub-intervals of equal length, fitted the model with $\delta=k/M$ for $k=0, \ldots, M$, and chose the best model from the $M+1$ candidate models according to a certain criterion.

\begin{figure}
	\centering
	\includegraphics[width=8cm]{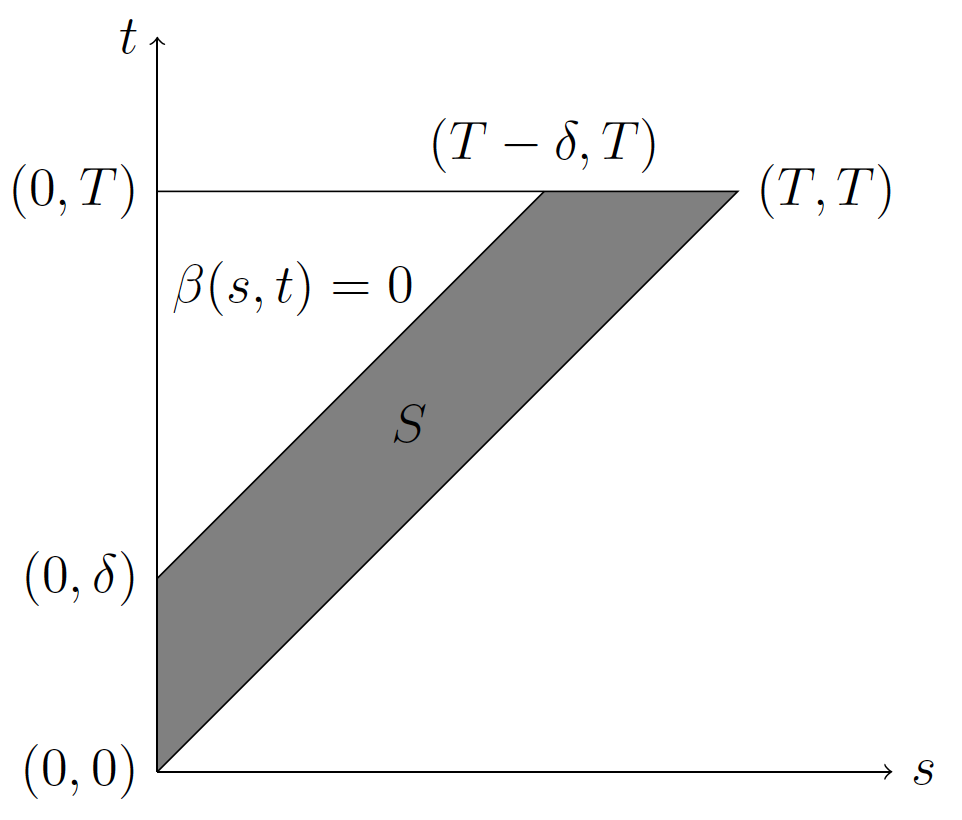}
	\caption{The support domain, $S$, of the coefficient function $\beta(s,t)$ as indicated by the gray area.}
	\label{fig:beta_region}
\end{figure}

In this work, we focus on the problem of estimating the historical lag and propose a tailored procedure for the simultaneous estimation of the coefficient function $\beta(s,t)$ and the lag $\delta$ in model~(\ref{mod:historical}). Considering the following ``full" model,
\begin{equation}
	\label{eq:mod}
	y_i(t) = \alpha(t) + 
	\int_0^t x_i(s)\beta(s,t) ds + \varepsilon_i(t),
	\mbox{ for } t \in [0, T].
\end{equation}
Figure~\ref{fig:beta_region} illustrates a scenario when the bivariate coefficient function $\beta(s, t)$ becomes zero over the triangular region defined by vertices $(0, \delta), (0, T), (T-\delta, T)$. This scenario corresponds to that $x(s)$ does not affect $y(t)$ for $s<t-\delta$. In other words, the full model (\ref{eq:mod}) is equivalent to the historical functional linear model (\ref{mod:historical}) when the support domain of the coefficient function $\beta(s,t)$ is a trapezoidal area $S$ defined by vertices $(0, 0), (T, T), (T-\delta, T), (0, \delta)$, given that the lag $\delta>0$. 

%

We propose a nested group bridge shrinkage method to estimate the historical lag $\delta$ and the coefficient function $\beta(s,t)$ with sparsity, which tackles the problem from a completely different perspective. Our method has two features. 
First of all, we utilize the triangular basis functions from the finite element method theory, which respects the non-rectangular support domain of the coefficient function $\beta(s,t)$. 
Such basis was also used by Malfait and Ramsay (2003) and Harezlak et al. (2007), as well as in spatial data analysis, where data distribute over irregularly shaped spatial domains with features like complex boundaries, strong concavities and interior holes (e.g. Ramsay, 2002; Sangalli, Ramsay and Ramsay, 2013). 
Then under model~(\ref{eq:mod}), we organize the basis coefficients in such a way that the nested group bridge penalty is able to shrink specifically the coefficient function $\beta(s,t)$ over the upper triangular region with vertices $(0,T)$, $(0,\delta)$ and $(T-\delta, T)$ towards zero. The group bridge shrinkage was originally proposed by Huang et al. (2009) for variable selection. Such penalty has been utilized by Wang and Kai (2015) for locally sparse estimation in nonparametric regression 
for a scalar-on-function historical functional linear model with B-spline basis function expansion.
%
%
The major advantage of our proposed approach is that we can estimate $\delta$ automatically without predetermining the candidate values of $\delta$. Our simulation studies show that our estimator of $\delta$ has a better finite sample performance than the conventional methods.

The rest of the paper is organized as follows. We introduce in Section~\ref{sec:method} the proposed estimation procedure as well as the computational details. 
Simulation studies are presented in Section~\ref{sec:simu}, and an application to speech production data in Section~\ref{sec:real}, and finally a summary in Section~\ref{sec:summary}.

\section{Method}
\label{sec:method}
To ease the notation, the intercept function $\alpha(t)$ can be dropped from model (\ref{eq:mod}) without loss of generality. Let $y_i^*(t) = y_i(t) -\overline{y}(t)$ and $x_i^*(t) = x_i(t) -\overline{x}(t)$ denote the pointwise centered response curves and predictor curves, respectively. A centered model without intercept is
\begin{equation}
\label{eq:mod0}
y^*_i(t) = \int_0^t x^*_i(s)\beta(s,t) ds + \varepsilon_i^*(t),
\quad t \in [0, T].
\end{equation}
Upon obtaining an estimate $\widehat{\beta}(s,t)$, the intercept function is then estimated as
$$
\widehat{\alpha}(t) = \overline{y}(t) - \int_{0}^{t} \overline{x}(s) \widehat{\beta}(s,t) ds.
$$
We drop the asterisk from model (\ref{eq:mod0}) from now on and focus the discussion on the estimation of the coefficient function $\beta(s,t)$ in the model: 
\begin{equation}
\label{eq:mod00}
y_i(t) = \int_0^t x_i(s)\beta(s,t) ds + \varepsilon_i(t),
\quad t \in [0, T].
\end{equation}

\subsection{Approximation with the Finite Element Method}
\label{sec:fem}
We propose to approximate the coefficient function $\beta(s,t)$ with the triangular basis, which originates from the finite element method and is widely used in the numerical solution for the boundary-value problems of partial differential equations. 
Noticing that the support of the coefficient function $\beta(s,t)$ is non-rectangular, approximation with a commonly used bivariate spline generated via tensor product would result in a jagged shape along the boundary $t=s$. Therefore, finite elements that can approximate arbitrary regions serve as a natural alternative.

Let $\phi_1(s,t), \ldots, \phi_K(s,t)$ denote the $K$ known triangular basis functions. The coefficient function $\beta(s,t)$ is approximated by the expansion, 
$$
\beta(s,t) \approx \sum_{k=1}^{K} b_k \phi_k(s,t).
$$
Plugging the above approximation into model (\ref{eq:mod0}), we obtain
\begin{eqnarray*}
	y_i(t) 
	&=& \sum_{k=1}^K b_k \int_{0}^{t} x_i(s) \phi_k(s,t) ds + \varepsilon_i^{'}(t) \\
	&=& \sum_{k=1}^K b_k \psi_{ik}(t) + \varepsilon_i^{'}(t),
\end{eqnarray*}
where $\psi_{ik}(t) = \int_{0}^{t} x_i(s) \phi_k(s,t) ds$ is known, $\varepsilon_i^{'}(t)$ includes both the residual $\varepsilon_i(t)$ and the approximation error.

\begin{figure}
	\centering
	\includegraphics[height=6cm]{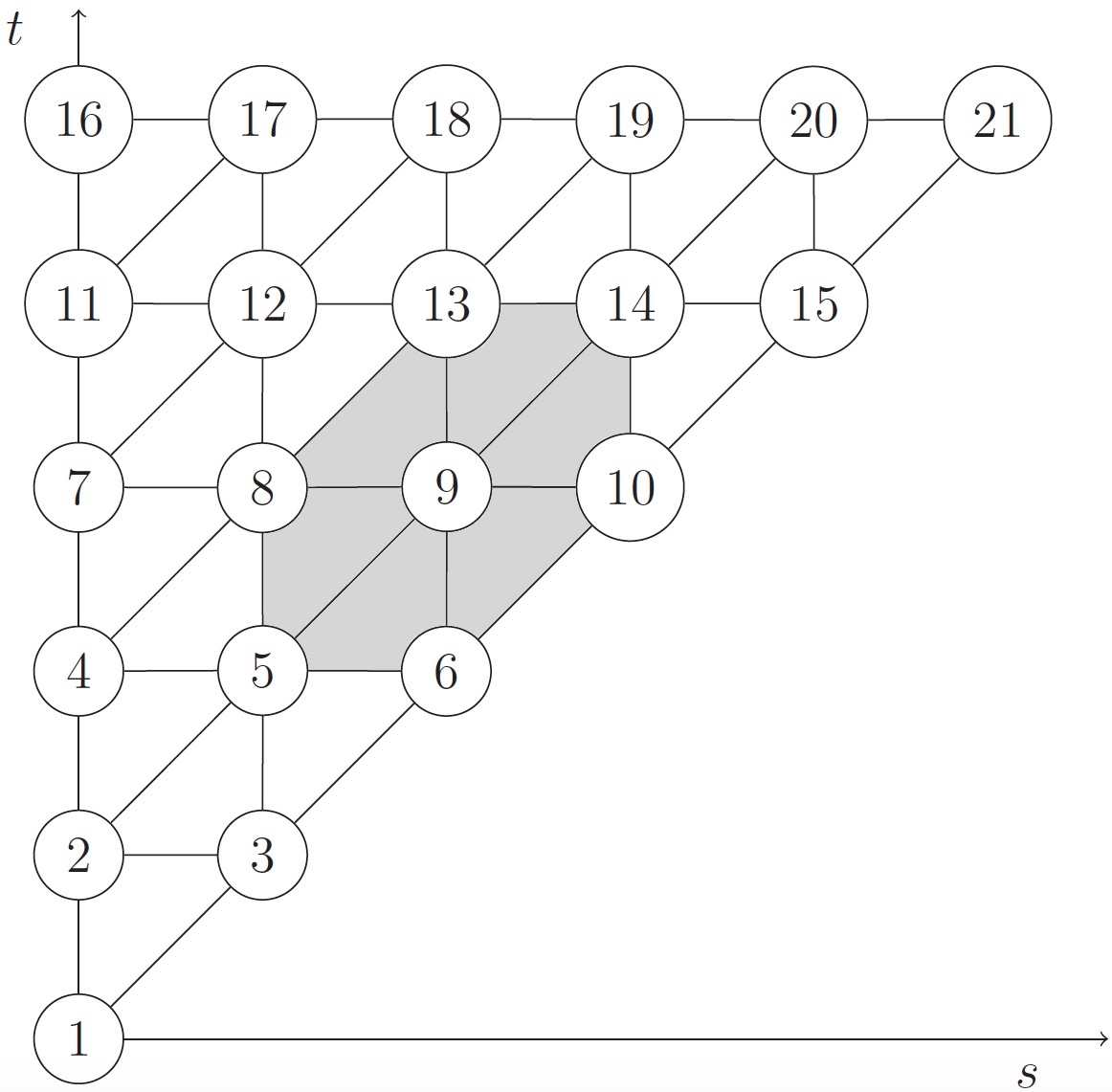}
	\hspace{0.2in}
	\includegraphics[height=6.5cm]{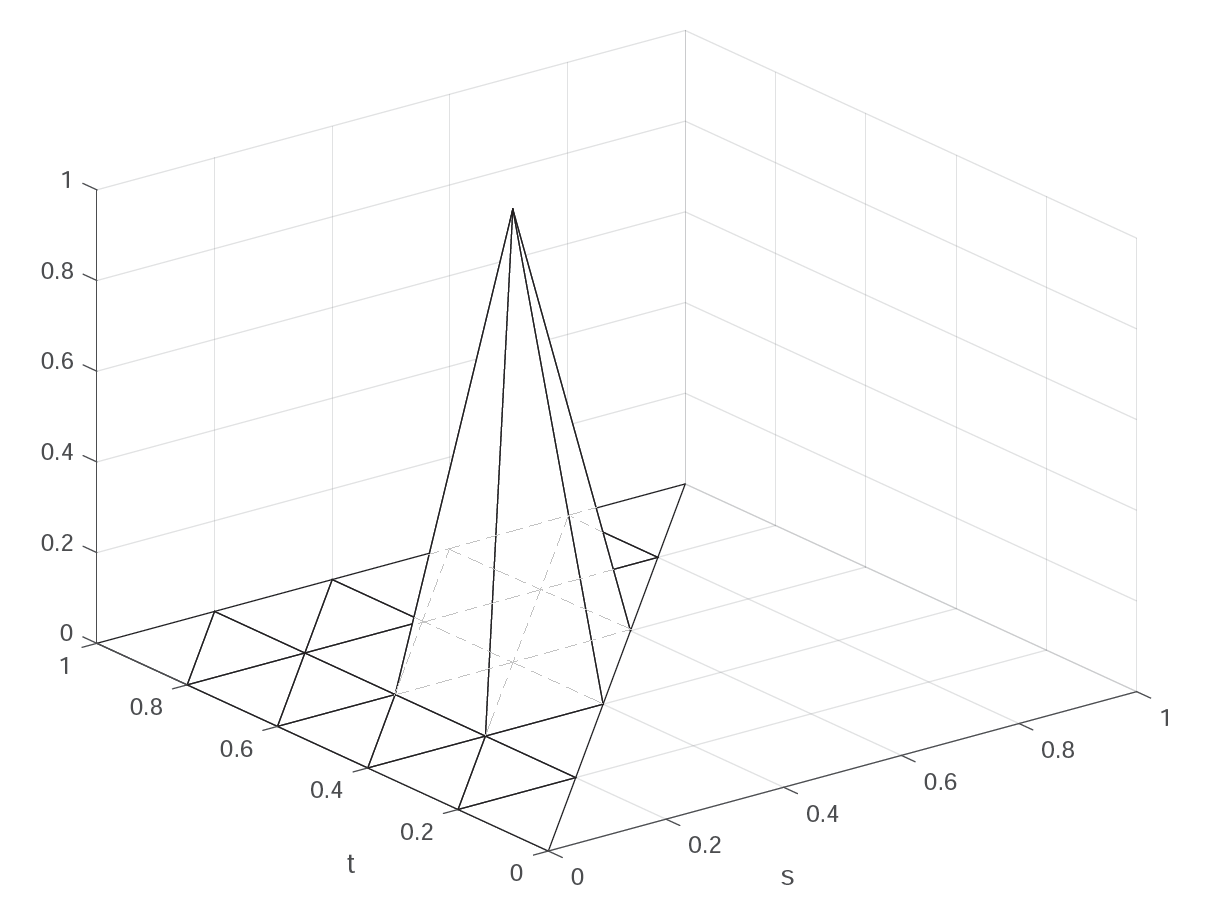}
	\caption{Illustration for the triangular basis system when $M=5$. There are 25 finite elements, 21 nodes, and 21 corresponding basis functions. Left: triangulation and indexation of the nodes. The gray hexagon indicates the support of the 9th basis function $\phi_9(s,t)$ which is shown on the right. Right: The basis function $\phi_9(s,t)$ corresponding to the 9th node at $(s,t)=(0.4,0.6)$ shown on the left. It is of a tent shape peaked at the 9th node.}
	\label{fig:tri_basis}
\end{figure}

Divide the interval $[0, T]$ on each axis 
into $M$ subintervals with equidistant nodes $0=t_0<t_1< \ldots <t_M=T$.
Further split each square into two triangles by the diagonal parallel to the line $t=s$. This divides the triangular region over which the coefficient function $\beta(s,t)$ is possibly non-zero into $M^2$ congruent triangles (i.e. the triangular elements, or finite elements) with $K=(M+1)(M+2)/2$ nodes. Each node corresponds to a basis function $\phi_k(s,t)$ and a coefficient $b_k$, both of which are indexed from bottom to top and row-wise from left to right. 
For example, the left panel in Figure~\ref{fig:tri_basis} shows the triangulation and indexation of the nodes when $M=5$. There are 25 triangular elements and 21 nodes corresponding to $b_k$ and $\phi_k(s,t)$ for $k=1,\ldots,21$.


Each basis function $\phi_k(s,t)$ has a compact support, defined over the hexagon centered at the $k$-th node. Basis functions of degree 1 are tent shaped, piecewise linear and continuous, with value 1 at node $k$ and value 0 at the boundary of the hexagon. For instance, the right panel in Figure~\ref{fig:tri_basis} shows a tent-shaped basis function corresponding to the 9th node at $(s,t)=(0.4,0.6)$. Though the $\phi_k(s,t)$'s are not orthogonal basis, their compact support property brings in certain computational advantage. We refer to Larson and Bengzon (2013) for a comprehensive introduction to the finite element basis system.

\subsection{The Nested Group Bridge Approach}
The coefficient function $\beta(s,t)$ in model~(\ref{eq:mod00}) is formally defined over the triangular region in Figure \ref{fig:beta_region} corresponding to the lag $\delta=T$, while we are trying to identify whether the upper left triangular region is zero or not. Therefore, a proper penalty should be able to shrink the upper left triangular region towards zero specifically, while respecting the nested group structure among the basis coefficients $b_k$'s.

We start by defining a sequence of nested triangular regions. Let $\Delta=T/M$ and $D_j$ denote the triangle with vertices $(0,T), (0, (j-1)\Delta), ((M+1-j)\Delta, T)$, for $j=1, \ldots, M$, and let $D_{M+1}=\{(0, T)\}$ contain a single point. Notice that $D_1 \supset \ldots \supset D_{M+1}$. In correspondence to these regions, we define a sequence of decreasing groups $A_1 \supset \ldots \supset A_{M+1}$, where $A_j$ consists of the indices of the nodes contained in triangle $D_j$ according to the node indexation described in Section~\ref{sec:fem}. 
Taking $M=3$ as an example, there are 10 nodes and the index sets are $A_1=\{7,4,8,2,5,9,1,3,6,10\}, A_2=\{7,4,8,2,5,9\}, A_3=\{7,4,8\}$, and $A_4=\{7\} $. 
Furthermore, we denote by $\bm{b}_{A_j}=\{b_k: k\in A_j\}$ the vector of basis coefficients whose indices belong to set $A_j$, and follow the conventional notations that $||\cdot||_{1}$ and $||\cdot||_2$ represent the $L_1$ and $L_2$ norms, respectively. 

We adapt the group bridge approach proposed by Huang et al. (2009) and propose to estimate the vector of basis coefficients $\bm{b}=(b_1, \ldots, b_K)$ for $\beta(s,t)$ in model~(\ref{eq:mod00}) by minimizing the penalized least squares criterion
\begin{equation}
\label{eq:ls_pen0}
\frac{1}{N}\int_{0}^{T} \sum_{i=1}^{N} 
\left\{ y_i(t) - \sum_{k=1}^{K} b_k \psi_{ik}(t) \right\}^2 dt 
+ \lambda \sum_{j=1}^{M+1} c_j||\bm{b}_{A_j}||^\gamma_1
+ \bm{b}^T \bm{R} \bm{b},
\end{equation}
with a fixed $\gamma \in (0,1)$, a nonnegative tuning parameter $\lambda$,  known weights $c_j$ to offset the effect of different dimensions of $A_j$, and a known smoothness penalty matrix $\bm{R}$ to be introduced shortly. Following Huang et al. (2009), 
a simple choice for the weights is $c_j \propto |A_j|^{1-\gamma}$.

The first term in criterion (\ref{eq:ls_pen0}) is the ordinary least squares taking into account the whole time course $[0, T]$. 
The second term is the so called \textit{nested group bridge} penalty, which originated from Huang et al. (2009) for simultaneous variable selection at both the group and within-group levels. 
Consider any two sets $A_j$ and $A_k$, with $j<k$. Due to the nested nature, the vector of coefficients $\bm{b}_{A_k}$ is always a subvector of $\bm{b}_{A_j}$, hence the coefficients corresponding to the nodes in region $D_k$ appear in more groups than the ones corresponding only to $D_j$. This suggests that the nested group bridge penalty shrinks more heavily the coefficients corresponding to regions in a closer proximity to $D_{M+1}$, as desired.
The last term imposes a discrete roughness penalty on the basis coefficients $\bm{b}$, extending the idea of Eilers and Marx (1996)
and Harezlak et al. (2007). Concerning the smoothness of the coefficient function $\beta(s,t)$, there are three interpretable directions in which we desire to control the changes in the adjacent coefficients, i.e., horizontal, vertical and diagonal directions, as shown in the left panel of Figure \ref{fig:tri_basis}. Along each direction, the rows of a penalty matrix correspond to the differences in the basis coefficients of adjacent nodes. The discrete smoothness penalty matrix is defined as
$$\bm{R}= \omega_H {\bm{D}_H}^T \bm{D}_H 
        + \omega_V {\bm{D}_V}^T \bm{D}_V
        + \omega_P {\bm{D}_P}^T \bm{D}_P, $$
which combines the three directions with nonnegative weights $\omega_H, \omega_V$, and $\omega_P$. An example for $M=3$ is as follows, where the columns of $\bm{D}_H$, $\bm{D}_V$ and $\bm{D}_P$ from left to right correspond to the basis coefficients $b_1, \ldots, b_{10}$,
\begin{center}
	$\bm{D}_H = \left[ {\begin{array}{*{20}{cccccccccc}}
		0&-1&1& 0& 0&0& 0& 0& 0&0\\
		0& 0&0&-1& 1&0& 0& 0& 0&0\\
		0& 0&0& 0&-1&1& 0& 0& 0&0\\
		0& 0&0& 0& 0&0&-1& 1& 0&0\\
		0& 0&0& 0& 0&0& 0&-1& 1&0\\
		0& 0&0& 0& 0&0& 0& 0&-1&1\\
		\end{array}} \right]$, 
	
	$\bm{D}_V = \left[ {\begin{array}{*{20}{cccccccccc}}
		-1& 1& 0& 0& 0& 0& 0& 0& 0&0\\
		0&-1& 0& 1& 0& 0& 0& 0& 0&0\\
		0& 0&-1& 0& 1& 0& 0& 0& 0&0\\
		0& 0& 0&-1& 0& 0& 1& 0& 0&0\\
		0& 0& 0& 0&-1& 0& 0& 1& 0&0\\
		0& 0& 0& 0& 0&-1& 0& 0& 1&0\\
		\end{array}} \right]$,
	
	$\bm{D}_P = \left[ {\begin{array}{*{20}{cccccccccc}}
		-1& 0& 1& 0& 0& 0& 0& 0& 0& 0\\
		0&-1& 0& 0& 1& 0& 0& 0& 0& 0\\
		0& 0&-1& 0& 0& 1& 0& 0& 0& 0\\
		0& 0& 0&-1& 0& 0& 0& 1& 0& 0\\
		0& 0& 0& 0&-1& 0& 0& 0& 1& 0\\
		0& 0& 0& 0& 0&-1& 0& 0& 0& 1\\
		\end{array}} \right]$.
\end{center}


\subsection{Computation}
The key difficulty in optimizing the objective function (\ref{eq:ls_pen0}) is due to its non-convexity. We can work on an equivalent constrained optimization problem, which is convex and easier to solve (Huang et al. 2009). The objective function (\ref{eq:ls_pen0}) involves an integration of the time dependent ordinary least squares criterion over time. Dividing the time course $[0,T]$ into $Q$ equally-spaced subintervals at time points $t_q$, $q=0, 1, \ldots, Q$, the integrated least squares criterion can be approximated by the finite sum as 
$$
\frac{T}{Q} \sum_{q=1}^Q \sum_{i=1}^{N}	 
\left\{ y_i(t_q) - \sum_{k=1}^{K} b_k \psi_{ik}(t_q) \right\}^2.
$$

With a sufficiently large $Q$, the above approximation with Riemann sums over a regular partition of the integral domain provides a satisfactory level of precision with acceptable computational load. 
Define the $Q$-by-$1$ vector $\bm{y}_i = (y_i(t_1), \ldots, y_i(t_Q)^T$, and the $Q$-by-$K$ matrix $\bm{\Psi}_i$ whose $(q,k)$-th element is $\psi_{ik}(t_q)$. By stacking the vectors $\bm{y}_i$ into a long vector $\bm{y}$ of length $NQ$ and the matrices $\bm{\Psi}_i$ into a tall matrix $\bm{\Psi}$ of size $NQ$ by $K$, the approximated integrated least squares criterion is rewritten in the matrix expression as
$$
\frac{T}{Q}\left(\bm{y} - \bm{\Psi}\bm{b}\right)^T
\left(\bm{y} - \bm{\Psi}\bm{b}\right).
$$
The objective function (\ref{eq:ls_pen0}) is then rewritten as the familiar penalized least squares
\begin{equation}
\label{eq:ls_pen1}
\frac{1}{N}\left(\bm{y} - \bm{\Psi}\bm{b}\right)^T
\left(\bm{y} - \bm{\Psi}\bm{b}\right)
+ \lambda \sum_{j=1}^{M+1} c_j||\bm{b}_{A_j}||^\gamma_1
+ \bm{b}^T \bm{R} \bm{b},
\end{equation}
where $\lambda$ differs from earlier notation by a constant factor $Q/T$.

Define a vector $\bm{\theta}=(\theta_1, \ldots, \theta_{M+1})$. Then $\widehat{\bm{b}}_n$ minimizes  criterion~(\ref{eq:ls_pen1}) if and only if $(\widehat{\bm{b}}_n, \widehat{\bm{\theta}})$ minimizes
\begin{equation}
\frac{1}{N}\left(\bm{y} - \bm{\Psi}\bm{b}\right)^T
\left(\bm{y} - \bm{\Psi}\bm{b}\right)
+ \sum_{j=1}^{M+1} \theta_j^{1-1/\gamma} c_j^{1/\gamma}||\bm{b}_{A_j}||_1 
+ \tau \sum_{j=1}^{M+1} \theta_j
+ \bm{b}^T \bm{R} \bm{b},
\label{eq:ls_pen_convex0}
\end{equation}
$$
s. t. \quad \theta_j \geq 0, j=1, \ldots, M+1,
$$
for $0<\gamma<1$ and $\tau=[\lambda \gamma^\gamma (1-\gamma)^{1-\gamma}]^{1/(1-\gamma)}$ (Huang et al. 2009). Notice that $|| \bm{b}_{A_j} ||_1 = \sum_{k \in A_j} |b_k|$ by definition of the $L_1$ norm, and that if $b_k$ appears in $A_j$ then it also appears in all $A_s$ with $s<j$. Thus, exchange the order of summation in the second term of  (\ref{eq:ls_pen_convex0}), rearrange and collect all the individual terms involving $b_k$, the objective function becomes
$$
\frac{1}{N}\left(\bm{y} - \bm{\Psi}\bm{b}\right)^T
\left(\bm{y} - \bm{\Psi}\bm{b}\right)
+ \sum_{k=1}^{K} g_k |b_k| 
+ \tau \sum_{j=1}^{M+1} \theta_j
+ \bm{b}^T \bm{R} \bm{b},
$$
$$
s. t. \quad \theta_j \geq 0, j=1, \ldots, M+1,
$$
where $g_k = \sum_{j=1}^{\ell(k)} \theta_j^{1-1/\gamma} c_j^{1/\gamma} $ with $\ell(k) = \max \{ j: b_k \in A_j \}$.  Intuitively, $\ell(k)$ is the number of times that coefficient $b_k$ appears in the nested group bridge penalty.


Define $\bm{G}$ as a $K \times K$ diagonal matrix with the $k$th diagonal element $(Ng_k)^{-1}$,
$\bm{b}^* = \bm{G}^{-1} \bm{b}$,
$\bm{\Psi}^*=[\bm{\Psi}^T, 
\sqrt{\omega_H N}\bm{D}_H^T, 
\sqrt{\omega_V N}\bm{D}_V^T, 
\sqrt{\omega_P N}\bm{D}_P^T]^T\bm{G}$,
and $\bm{y}^*=(\bm{y}^T, \bm{0}^T)^T$ of proper length. 
Finally, the objective function (\ref{eq:ls_pen_convex0}) is expressed as
$$
\frac{1}{N} 
\left\{ 
\left( \bm{y}^* - \bm{\Psi}^*\bm{b}^* \right)^T
\left(\bm{y}^* - \bm{\Psi}^*\bm{b}^*\right)
+ \sum_{k=1}^{K} |b_k^*|
\right\}
+ \tau \sum_{j=1}^{M+1} \theta_j,
$$
$$
s. t. \quad \theta_j \geq 0, j=1, \ldots, M+1,
$$
where $b_k^*$ is the $k$th element of vector $\bm{b}^*$. The following iterative algorithm is used to compute $\widehat{\bm{b}}_n$.
\begin{itemize}
		\item \textit{Step 1}: Obtain an initial estimate $\bm{b}^{(0)}$ with ordinary least squares.

		\item \textit{Step 2}: At each iteration $s$, update $\bm{\theta}$ based on ${\bm{b}}^{(s-1)}$ as
		$$ \theta_j^{(s)} = c_j 
		\left( \frac{1-\gamma}{\tau \gamma} \right)^\gamma 
		|| \bm{b}_{A_j}^{(s-1)}||_1^\gamma, 
		\quad j=1, \ldots, M+1, $$
		and correspondingly
		$$ g_k^{(s)} = \sum_{j=1}^{\ell(k)}
		\left( \theta_j^{(s)} \right) ^{1-1/\gamma} c_j^{1/\gamma},
		\quad k=1, \ldots, K,$$
		$$\bm{G}^{(s)} = 
		N^{-1} \mbox{diag}
		\left( 1/g_1^{(s)}, \ldots, 1/g_K^{(s)} \right),$$
		$$ \bm{\Psi}^{*(s)} = 
		[\bm{\Psi}^T, 
		\sqrt{\omega_H N}\bm{D}_H^T, 
		\sqrt{\omega_V N}\bm{D}_V^T, 
		\sqrt{\omega_P N}\bm{D}_P^T]^T
		\bm{G}^{(s)}.$$
	
		\item \textit{Step 3}: At each iteration $s$, update $\bm{b}$ based on $\bm{\theta}^{(s)}$ by recognizing this is a LASSO problem
		$$ \bm{b}^{(s)} = \bm{G}^{(s)} \mbox{argmin}_{\bm{b}^*}
		\left\{ 
		\left( \bm{y} - \bm{\Psi}^{*(s)}\bm{b}^* \right)^T
		\left(\bm{y} - \bm{\Psi}^{*(s)}\bm{b}^*\right)
		+ \sum_{k=1}^{K} |b_k^*|
		\right\}.
		$$
	
		\item Repeat \textit{Step 2} and \textit{Step 3} until convergence.
\end{itemize}

Once obtaining $\widehat{\bm{b}}_n$, the estimators for $\beta(s,t)$ and $\delta$ are defined as 
$$
\widehat{\beta}_n(s,t) = \sum_{k=1}^{K} \widehat{b}_{n, k} \phi_k(s,t), 
\quad
\widehat{\delta}_n = \frac{T}{M}
\min_j \{1\leq j \leq M: \widehat{\beta}_n(s,t) =0 \mbox{ on } D_j\}.
$$
After obtaining $\widehat{\delta}_n$, we then refine the estimation for $\beta(s,t)$ by minimizing criterion (\ref{eq:ls_pen1}) with $\lambda=0$, i.e. excluding the nested grouped bridge penalty.

When implementing the algorithm, there are two types of tuning parameters, i.e. the shrinkage parameter $\lambda$ (or equivalently $\tau$) and the smoothness parameters as one type, and the number of grids on the time interval $[0, T]$ as the other type. Since the precision of $\widehat{\delta}_n$ obviously depends on $M$, we desire a relatively large $M$ in order to achieve a reasonably good estimate of the historical lag $\delta$, and at the same time to capture enough local features of the coefficient function $\beta(s,t)$. This is also consistent with the common strategy when applying penalized least squares approach, where a relatively large number of nodes is preferred and potential overfitting caused by such a choice would be offset via the sparsity and roughness penalty. The shrinkage parameter and smoothness parameters can be chosen via, for example, the Bayesian information criterion (BIC). The effective degrees of freedom for given $\lambda$ and $\bm{\kappa}=(\omega_H, \omega_V, \omega_P)$ can be approximated by
$$
df(\lambda, \bm{\kappa}) = \mbox{trace}(\bm{\Psi}_s ( \bm{\Psi}_s^T \bm{\Psi}_s + N \bm{R}_s )^{-1} \bm{\Psi}_s^T ),
$$
where $\bm{\Psi}_s$ consists of the columns of $\bm{\Psi}$ corresponding to the nonzero coefficients in $\widehat{\bm{b}}_n$, and $\bm{R}_s$ is obtained in the same way by properly selecting the columns of $\bm{D}_H$, $\bm{D}_V$ and $\bm{D}_P$.
%
%
%
The BIC is then approximated by
$$
BIC(\lambda, \bm{\kappa}) = N \log( || \bm{y} - \bm{\Psi} \widehat{\bm{b}}_n(\lambda, \bm{\kappa})||_2^2 /N ) + \log(N) df(\lambda, \bm{\kappa}),
$$
and the tuning parameters are selected as the minimizer of $BIC(\lambda, \bm{\kappa})$.

%
%
%
%

\section{Analysis of Real Data}
\label{sec:real}
In this analysis, we apply the proposed nested group bridge approach to data from a speech production experiment. It is known that there are over 100 muscles that must be controlled centrally during speech production,  such as the muscles of the thoracic and abdominal walls, the neck and face, and the oral cavity, etc. 
The timing of activation of different muscle groups is a central issue in anatomical and physiological research of speech. Due to the fact that muscle contractions generate electrochemical changes, the noninvasive electromyography (EMG) method is used to collect data on muscle activation. 
With electrodes attached to the skin over the muscle, potentials as a result of muscle stimulation can be picked up. 
In this experiment, a subject said the syllable ``bob'' 32 times. By Newton's second law, the accelerations of the center of the lower lip reflect the force applied to tissue by muscle contraction, which was recorded as the response curves. Ramsay and Silverman (2002) provided a comprehensive introduction of the background knowledge about this physiology study.

\begin{figure}
	\centering
	\includegraphics[height=5cm, width=14cm]{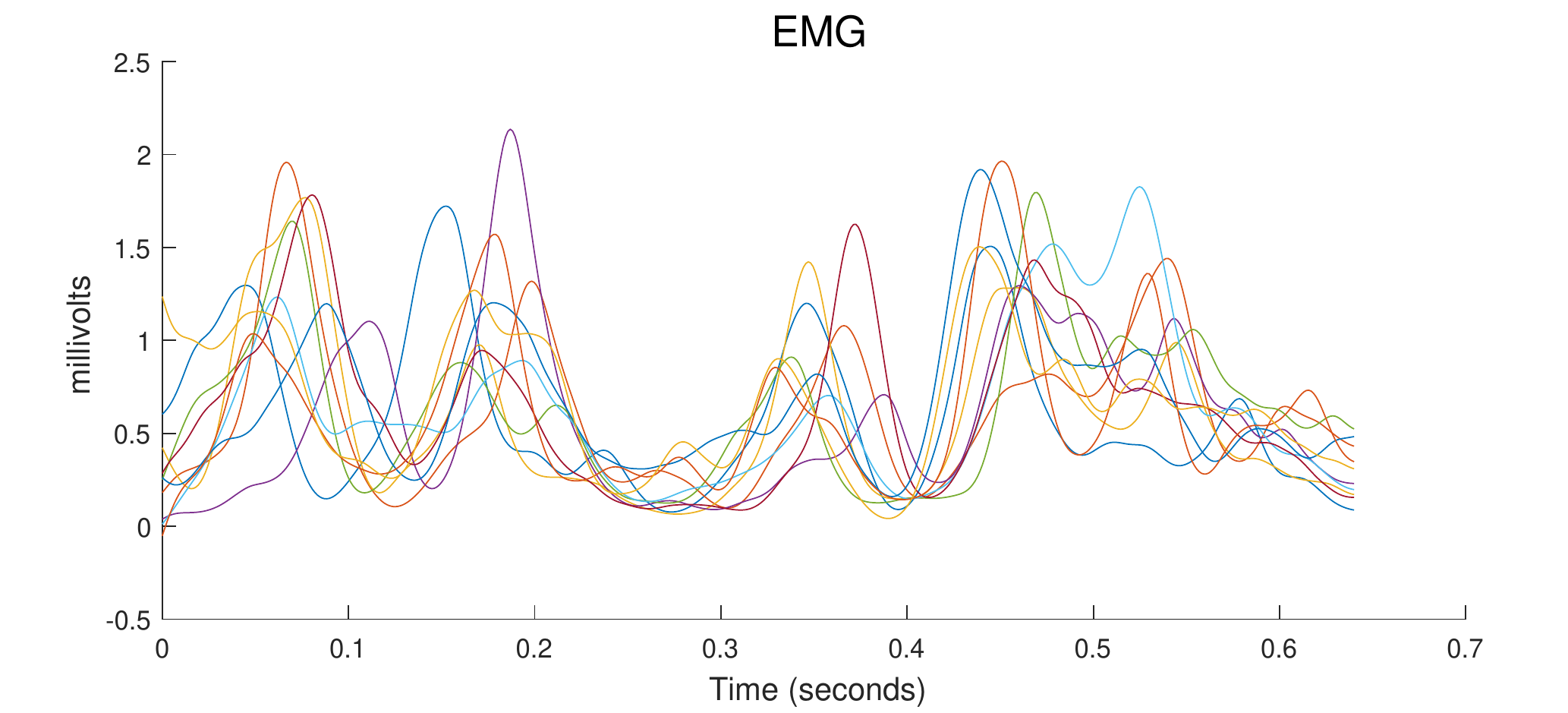} \\
	\includegraphics[height=5cm, width=14cm]{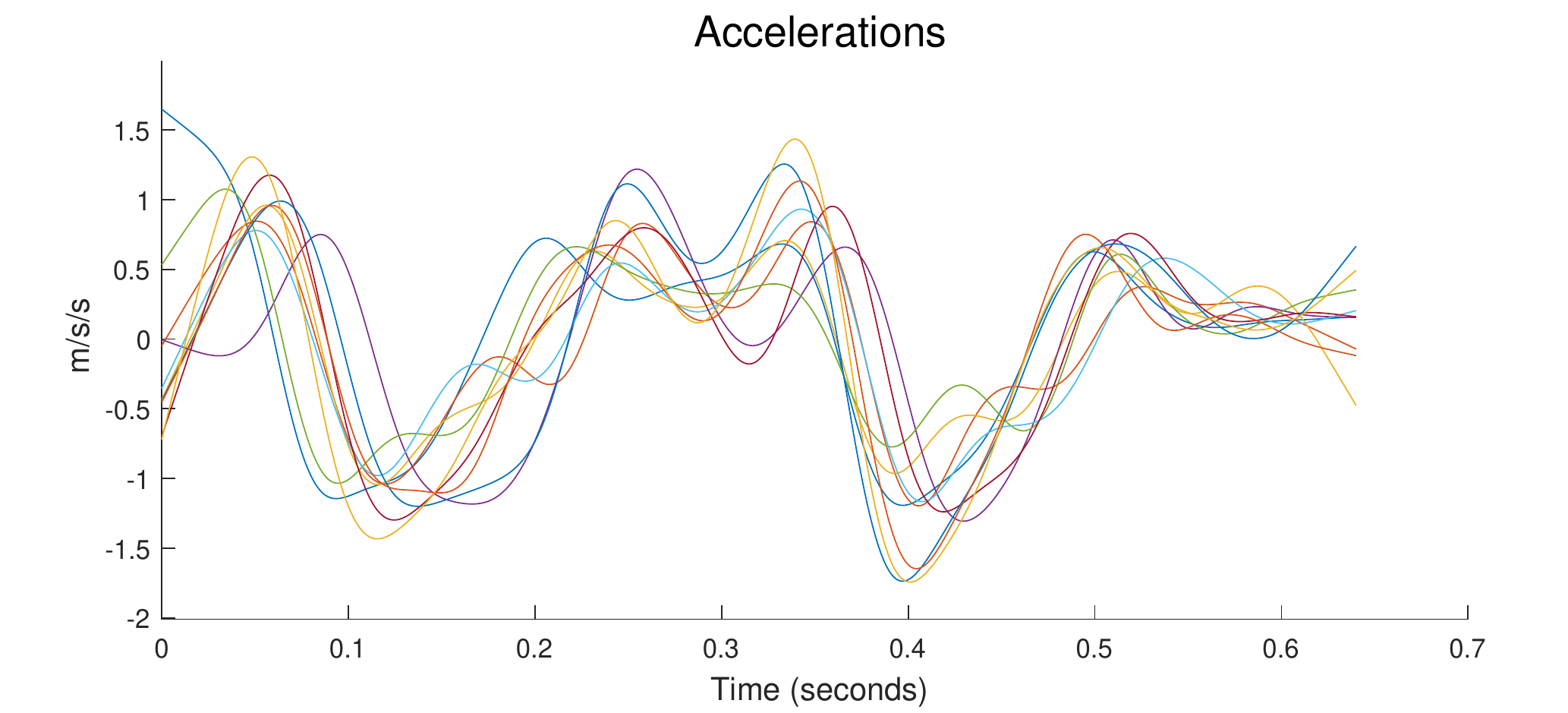}
	\caption{A random subset of 10 pairs of EMG and lip acceleration curves. The observation time range is $[0, 0.64]$. Top: the EMG activities associated with the depressor labii inferior muscle. Bottom: the acceleration of the center of the lower lip. This figure appears in color in the electronic version of this article.}
	\label{fig:xy_curves}
\end{figure}

A random subset of 10 smoothed observations for EMG recordings (top) and corresponding lower lip accelerations (bottom) is presented in Figure~\ref{fig:xy_curves}, where the time range is $[0, 0.64]$.
Considering that the current lip acceleration is very likely a result of recent muscle movement, this motivates us to fit the historical functional linear model (\ref{mod:historical}) with an unknown lag. The EMG recordings are the functional covariates and the lip accelerations are the functional response. The goal of this analysis is to explore the association between EMG recording and lip acceleration, and identify a historical lag if there is any. 
To apply the nested group bridge approach, we divides the time range [0,0.64] into 20 subintervals with equidistant nodes, i.e. $M=20$. This leads to $K=231$ nodes and triangular basis functions. 
By minimizing criterion (\ref{eq:ls_pen0}) as described in Section~\ref{sec:method}, we obtain the estimates for both the regression coefficient and the historical lag.

\begin{figure}
	\centering
	\includegraphics[height=8cm, width=12cm]{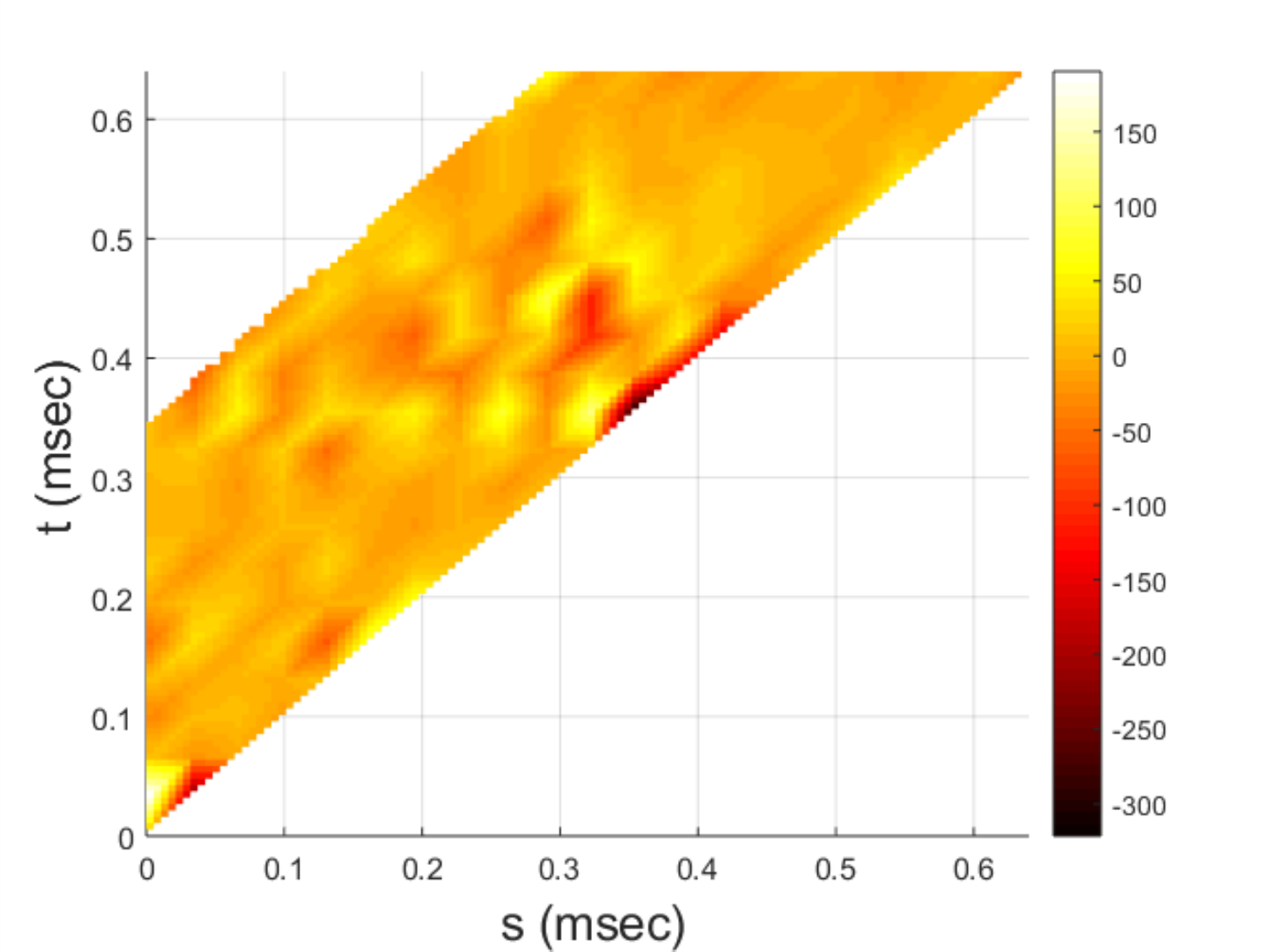}
	\caption{The estimated regression coefficient function $\widehat{\beta}(s,t)$ for a historical lag of 0.352 estimated from the data. This figure appears in color in the electronic version of this article.}
	\label{fig:beta_hat}
\end{figure}

The estimated historical lag is $\widehat{\delta}=0.352$ seconds, indicating there is a historical effect of muscle activation on the lip movement. A 95\% confidence interval for the historical lag $\delta$ is $(0.348, 0.406)$, constructed via the bootstrap method by re-sampling the residuals and then re-estimating the model. 
Figure~\ref{fig:beta_hat} shows the corresponding estimate of the bivariate coefficient function ${\beta}(s,t)$. Dividing $\widehat{\beta}(s,t)$ pointwise by its standard error obtained via the bootstrap method, the plot of the test statistics looks very similar to Figure~\ref{fig:beta_hat}, with slight difference in the magnitude of the values. We observe that most of the coefficients are significant, using a two-sided $z$-test at 5\% significance level. The estimate of $\beta(s,t)$ has extraordinarily large values along the diagonal direction of two regions, $(s,t)\in (0, 0.06)\times(0, 0.06)$ and $(s,t)\in (0.33, 0.44)\times(0.33, 0.44)$. This is as expected, corresponding to the two productions of the \verb|/b/| syllable when the lip is closed and the muscle activation is most influential. The width of the diagonal band with large values varies approximately from 50 to 60 milliseconds, which corresponds to the delay for a neural signal to be transduced into muscle contraction. And peaks at larger lags suggest possible covariation of EMG and lip movement that requires solid physiological knowledge for interpretation.


\section{Simulation Studies}
\label{sec:simu}
\begin{figure}
	\centering
	\begin{tabular}{lll}
		\includegraphics[width=5.5cm]{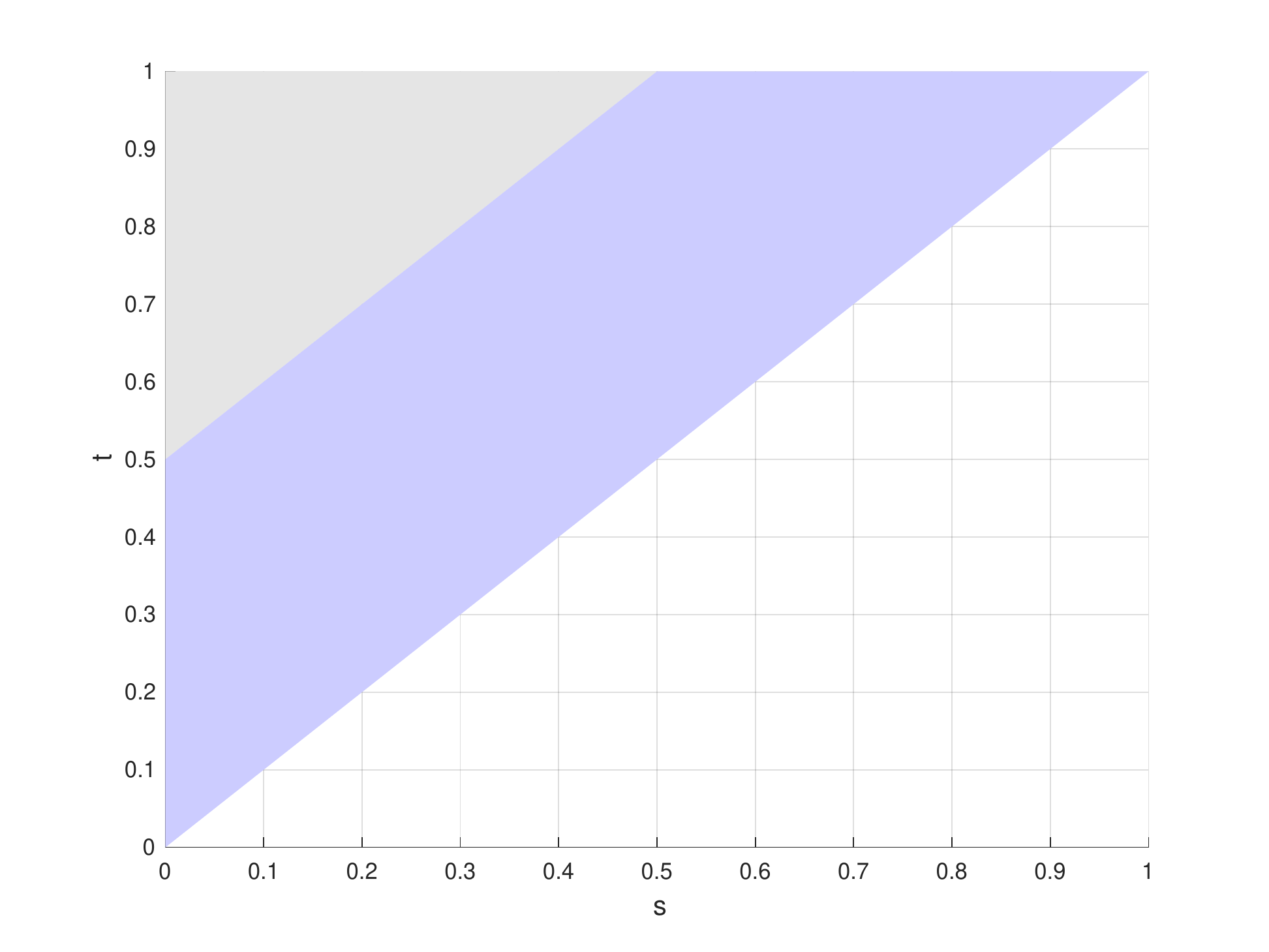} &
		\includegraphics[width=5.5cm]{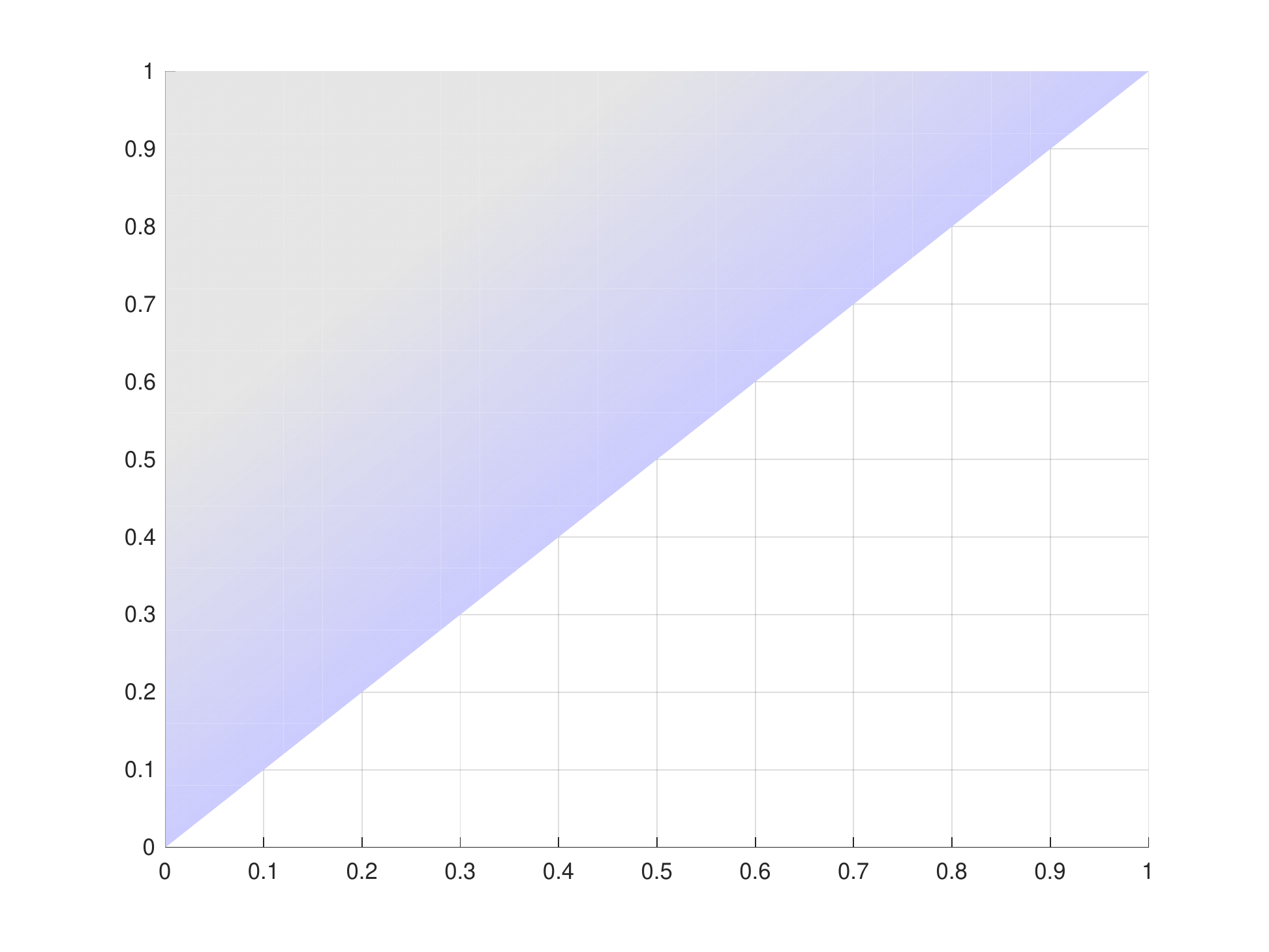} &
		\includegraphics[width=5.5cm]{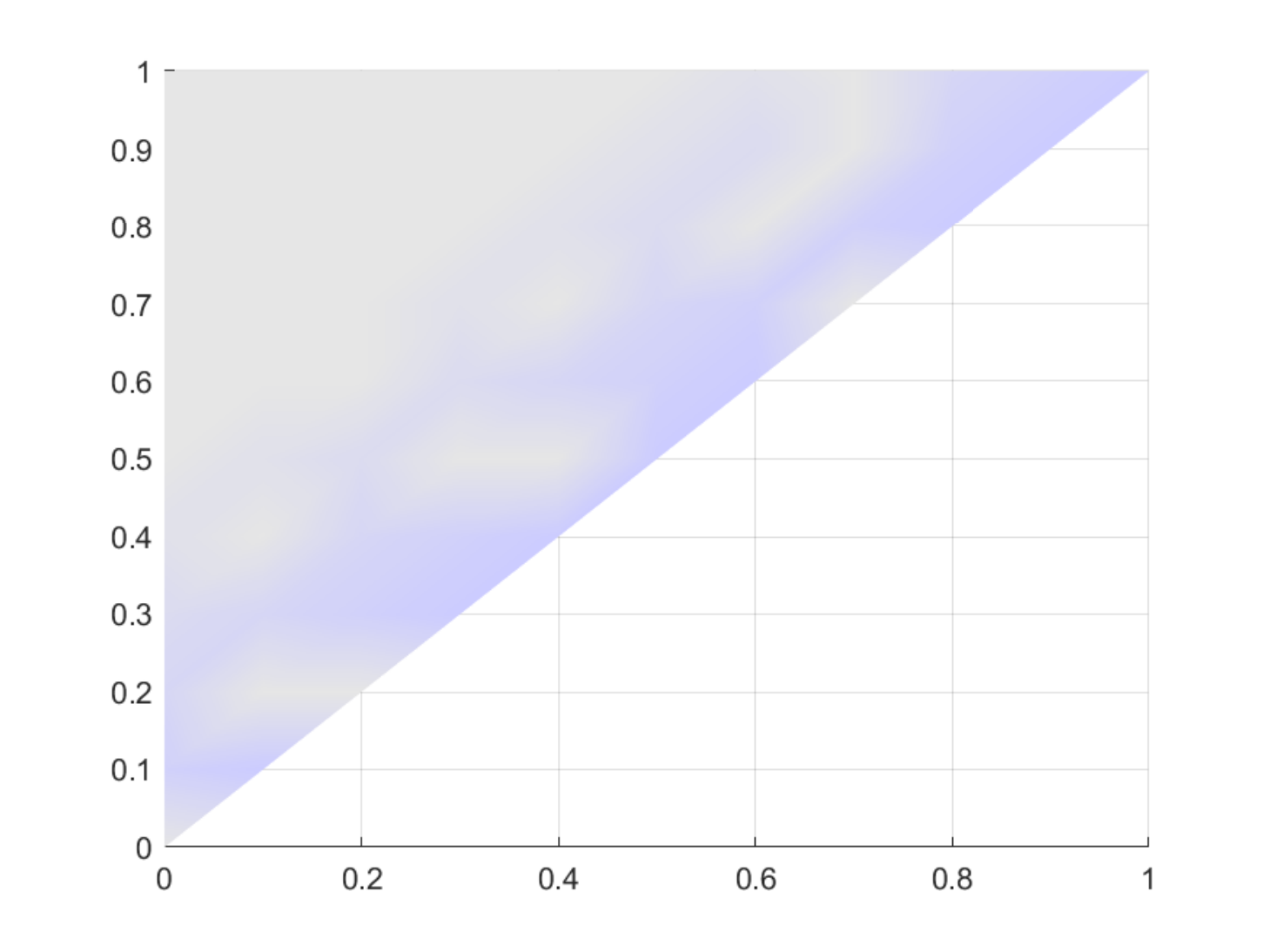} 
		\vspace{10pt}
	\end{tabular}
	\caption{The true coefficient function $\beta(s,t)$ used in simulation scenarios. From left to right are the truth for Scenario 1, 2, and 3, respectively, with increasing difficulty in the estimation. This figure appears in color in the electronic version of this article.}
	\label{fig:scenarios}
\end{figure}

Simulation studies are conducted to evaluate the proposed estimation procedure with a nested group bridge penalty (NGB) in comparison with the penalized approach by Harezlak et al. (2007) based on linear model approximation (PLMA). 
Set $T=1$. Let $S$ denote the trapezoidal support with vertices $(0,0)$, $(1,1)$, $(1-\delta, 1)$, $(0, \delta)$ as shown in Figure~\ref{fig:beta_region}, and $I(\cdot)$ the indicator function. We have considered the following three scenarios, and the true coefficient function $\beta(s,t)$ is shown in Figure~\ref{fig:scenarios}.

\noindent{\bf Scenario 1.} $\beta(s,t)=10 I\{(s,t) \in S_{\varepsilon} \} + 10(\frac{\delta}{\varepsilon} + \frac{s}{\varepsilon}  - \frac{t}{\varepsilon} ) I\{(s,t) \in S\backslash S_{\varepsilon} \}$, where $S_{\varepsilon}$ is the trapezoid with vertices $(0,0)$, $(1,1)$, $(1-\delta+\varepsilon, 1)$, $(0, \delta-\varepsilon)$. The coefficient function $\beta(s,t)$ is constant over the region $S_{\varepsilon}$, linear over $S\backslash S_{\varepsilon}$ and vanishes at the line $t=s+\delta$.

\noindent{\bf Scenario 2.} $\beta(s,t)= 10(1 + \frac{s}{\delta} - \frac{t}{\delta} ) \times I\{(s,t)\in S\}$. The coefficient function $\beta(s,t)$ is linear over its support $S$, with maximum along the line $t=s$ and vanishes at the line $t=s+\delta$.

\noindent{\bf Scenario 3.} Some ``holes'' are created randomly on the coefficient function $\beta(s,t)$ of Scenario~2, such that $\beta(s,t)$ can vanish inside $S_{\varepsilon}$. 


In Scenario 1, we take a small value of $\varepsilon=0.05$, which leads to a sharp drop in the coefficient function $\beta(s,t)$ towards zero when going outside from the region $S_{\varepsilon}$ toward the line $t=s+\delta$. We consider this scenario as the easiest situation to determine the historical lag $\delta$ due to the sharp change, and expect both approaches to perform well and similarly. Scenario 2 is slightly more difficult, since the boundary $t=s+\delta$ becomes more blurred than that in Scenario 1 when data are contaminated with errors. 
Scenario 3 adds further difficulty with irregularly shaped coefficient function $\beta(s,t)$ and admits the more general assumption that the dependence of response on the historical predictor can vary over time. 
For all scenarios, the observed covariate curves in the real data example presented in Section~\ref{sec:real} are rescaled to time interval $[0, 1]$ and are used along with the true coefficient function $\beta(s,t)$ to generate the mean response curves, possibly to mimic real situations. A random subset of 10 covariate curves are shown in the top panel of Figure~\ref{fig:xy_curves}. We refer to Section~\ref{sec:real} for details about the example. The errors are additive and generated pointwise from a $N(0, 0.5^2)$ distribution.

For the proposed NGB approach, we set  $\gamma=0.5$ and use the ordinary least squares estimate by Malfait and Ramsay (2003) as an initial value for the algorithm. We choose $c_j=|A_j|^{1-\gamma} / || \bm{b}_{A_j}^{(0)} ||_2^\gamma$ similarly to the idea of adaptive LASSO (Zou, 2006).
For each scenario, 100 independent replications are simulated. The results are summarized in Table~\ref{tb:simu}. Concerning the estimated root mean squared error of $\widehat{\delta}$, the proposed NGB estimator outperforms the PLMA estimator in all three scenarios, with a much smaller bias and comparable standard deviation. While the PLMA estimator performs well in Scenario 1 with a small bias and a reasonably small standard deviation, the performance deteriorates badly in more difficult scenarios and the estimator exhibits a tendency to underestimate $\delta$. 
To assess the accuracy of the estimated bivariate coefficient function, we calculate the square root of the mean integrated squared errors (RISEs) over a meshgrid, also presented in Table~\ref{tb:simu}. The NGB and PLMA approaches do not shown discernible difference in terms of the estimated RISE for $\widehat{\beta}(s,t)$. 


\begin{table}
	\centering
	 \def\~{\hphantom{0}}
 \begin{minipage}{175mm}
	\caption{Summary of the 100 simulation replications for each scenario, including the root mean squared error (RMSE), percent bias (\%bias) and standard deviation (SD) of $\widehat{\delta}$, and the square root of the mean integrated squared errors (RISE) of $\widehat{\beta}(s,t)$ along with the corresponding standard deviation in parentheses. The two methods implemented are the proposed nested group bridge approach (NGB) and the penalized linear model approximation approach (PLMA).}
	\label{tb:simu}
	\begin{tabular}{ccccrrccccrr}
		\hline
		\hline
       &  \multicolumn{2}{c}{RMSE of $\widehat{\delta}$} & & \multicolumn{2}{c}{\%bias of $\widehat{\delta}$} & & \multicolumn{2}{c}{SD of $\widehat{\delta}$} & & \multicolumn{2}{c}{RISE of $\widehat{\beta}(s,t)$} \\
	   \cline{2-3} \cline{5-6} \cline{8-9} \cline{11-12} \vspace{-10pt}\\  
	   & NGB & PLMA  & & \multicolumn{1}{c}{NGB} & \multicolumn{1}{c}{PLMA}  & & NGB & PLMA  & & \multicolumn{1}{c}{NGB} & \multicolumn{1}{c}{PLMA}  \\ \hline
	   Scenario 1 & 0.0173         & 0.0206      & & 	  1.2 & 1.7 & & 0.0162  &  0.0188   & &  1.29 (0.25)   &  0.83 (0.35)   
	   \\
	   Scenario 2 & 0.0480         & 0.1587       & &   -5.4  & -31.5 & & 0.0396 & 0.0192  & & 69.86 (7.04)  & 69.33 (7.54)  \\
	   Scenario 3 & 0.0548         & 0.2426    & &  -9.2   & -48.3 &  & 0.0297 &   0.0235 & &  8.74 (1.33)  &  9.75 (1.34)  \\
	   \hline
	\end{tabular}
\end{minipage}
\end{table}


%

\section{Summary}
\label{sec:summary}
We have considered in this article the estimation of the historical functional linear model~(\ref{mod:historical}) with an unknown forward lag. We propose a nested group bridge approach, tailored for the simultaneous estimation of the historical lag and the regression coefficient function. The nested group bridge penalty is able to shrink not only a group of coefficient corresponding to the designated area towards zero, but also individual coefficients corresponding to the remaining area towards zero. We adopt the triangular basis from the finite element method in order to conform naturally to the non-rectangular domain of the regression coefficient function. The triangular basis system is computationally efficient in the sense that the compact support of the basis functions leads to a sparse design matrix.



Under the historical functional linear model (\ref{mod:historical}), we assume that the historical lag $\delta$ is independent of $t$. It may be of interest to describe more precisely the historical effect of the functional covariate on the functional response via a time-dependent lag $\delta(t)$. Another point of interest is the alternative choice of the smoothness penalty. The discrete penalty originated from P-spline (Eilers and Marx, 1996) involves a large matrix and its inversion. It is of interest to explore other possibilities. We leave the above mentioned points for future work.



\nocite{*}
\bibliographystyle{acm}

\begin{thebibliography}{}
	\bibitem{ } Asencio, M., Hooker, G., and Gao, H. O. (2014). Functional convolution models. \textit{Statistical Modelling: An International Journal}, \textbf{14(4)}, 315--335.
	
	\bibitem{ }	Besse, P. C., and Cardot, H. (1996). Approximation spline de la pr\'{e}vision d'un processus fonctionnel autor\'{e}gressif d'ordre 1. \textit{Canadian Journal of Statistics}, \textbf{24(4)}, 467--487.
	
	\bibitem{ } Brockhaus, S., Melcher, M., Leisch, F., and Greven, S. (2017). Boosting flexible functional regression models with a high number of functional historical effects. \textit{Statistics and Computing}, \textbf{27(4)}, 913--926.
	
	\bibitem{ }	Chen, H., and Wang, Y. (2011). A penalized spline approach to functional mixed effects model analysis. \textit{Biometrics}, \textbf{67(3)}, 861--870. 
	
	\bibitem{ } Eilers, P. H., and Marx, B. D. (1996). Flexible smoothing with B-splines and penalties. \textit{Statistical Science}, \textbf{11(2)}, 89--102.

	\bibitem{ }	Fan, J., and Zhang, J.-T. (2000). Two-step estimation of functional linear models with applications to longitudinal data. \textit{Journal of the Royal Statistical Society, Series B}, \textbf{62(2)}, 303--322. 

	\bibitem{ } Ferraty, F., and Vieu, P. (2006). \textit{Nonparametric Functional Data Analysis: Theory and Practice}. New York: Springer.

	\bibitem{ }	Greven, S., and Scheipl, F. (2017). A general framework for functional regression modelling. \textit{Statistical Modelling: An International Journal}, \textbf{17(1--2)}, 1--35.
	
	\bibitem{ }	Harezlak, J., Coull, B. A., Laird, N. M., Magari, S. R., and Christiani, D. C. (2007). Penalized solutions to functional regression problems. \textit{Computational Statistics and Data Analysis}, \textbf{51(10)}, 4911--4925. 
	
	\bibitem{ } Hastie, T., and Tibshirani, R. (1993). Varying-coefficient models. \textit{Journal of the Royal Statistical Society, Series B}, \textbf{55(4)}, 757--779. 
	
	\bibitem{ } Hsing, T., and Eubank, R. (2015). \textit{Theoretical Foundations of Functional Data Analysis, with an Introduction to Linear Operators}. New York: John Wiley \& Sons.

	\bibitem{ } Huang, J., Ma, S., Xie, H., and Zhang, C.-H. (2009). A group bridge approach for variable selection. \textit{Biometrika}, \textbf{96(2)}, 339--355.

	\bibitem{ }	Ivanescu, A. E., Staicu, A.-M., Scheipl, F., and Greven, S. (2015). Penalized function-on-function regression. \textit{Computational Statistics}, \textbf{30(2)}, 539--568. 

	\bibitem{ } Kim, K., \c{S}ent\"{u}rk, D., and Li, R. (2011). Recent history functional linear models for sparse longitudinal data. \textit{Journal of Statistical Planning and Inference}, \textbf{141(4)}, 1554--1566. 
	
	\bibitem{ } Kokoszka, P., and Reimherr, M. (2017). \textit{Introduction to Functional Data Analysis}. New York: CRC Press.

	\bibitem{ } Larson, M. G., and Bengzon, F. (2013). \textit{The Finite Element Method: Theory, Implementation, and Applications}. New York: Springer.
	
	\bibitem{ } Malfait, N., and Ramsay, J. O. (2003). The historical functional linear model. \textit{Canadian Journal of Statistics}, \textbf{31(2)}, 115--128.
	
	\bibitem{ } Morris, J. S. (2015). Functional regression. \textit{Annual Review of Statistics and Its Application}, \textbf{2(1)}, 321--359. 

	\bibitem{ } M\"{u}ller, H.-G., and Zhang, Y. (2005). Time-varying functional regression for predicting remaining lifetime distributions from longitudinal trajectories. \textit{Biometrics}, \textbf{61(4)}, 1064--1075. 
		
	\bibitem{ }  Ramsay, J. O., and Dalzell, C. J. (1991). Some tools for functional data analysis. \textit{Journal of the Royal Statistical Society, Series B}, \textbf{53(3)}, 539--561. 
	
	\bibitem{ } Ramsay, J. O., and Silverman, B. W. (2002). \textit{Applied Functional Data Analysis: Methods and Case Studies}. New York: Springer. 
	
	\bibitem{ } Ramsay, J. O., and Silverman, B. W. (2005). \textit{Functional Data Analysis (2nd ed)}. New York: Springer.

	\bibitem{ } Ramsay, T. (2002). Spline smoothing over difficult regions. \textit{Journal of the Royal Statistical Society, Series B}, \textbf{64(2)}, 307--319. 
	
	\bibitem{ } Sangalli, L. M., Ramsay, J. O., and Ramsay, T. O. (2013). Spatial spline regression models. \textit{Journal of the Royal Statistical Society, Series B}, \textbf{75(4)}, 681--703.
	
	\bibitem{ } \c{S}ent\"{u}rk, D., and M\"{u}ller, H.-G. (2008). Generalized varying coefficient models for longitudinal data. \textit{Biometrika}, \textbf{95(3)}, 653--666. 

	\bibitem{ }	\c{S}ent\"{u}rk, D., and M\"{u}ller, H.-G. (2010). Functional varying coefficient models for longitudinal data. \textit{Journal of the American Statistical Association}, \textbf{105(491)}, 1256--1264.
	
	\bibitem{ } Wang, H., and Kai, B. (2015). Functional sparsity: global versus local. \textit{Statistica Sinica}, \textbf{25(4)}, 1337--1354. 
		
	\bibitem{ } Wang, J.-L., Chiou, J.-M., and M\"{u}ller, H.-G. (2016). Functional data analysis. \textit{Annual Review of Statistics and Its Application}, \textbf{3(1)}, 257--295.

	\bibitem{ } Wu, C. O., Chiang, C.-T., and Hoover, D. R. (1998). Asymptotic confidence regions for kernel smoothing of a varying-coefficient model with longitudinal data. \textit{Journal of the American Statistical Association}, \textbf{93(444)}, 1388--1402.
	
	\bibitem{ } Wu, H., and Liang, H. (2004). Backfitting random varying-coefficient models with time-dependent smoothing covariates. \textit{Scandinavian Journal of Statistics}, \textbf{31(1)}, 3--19. 

	\bibitem{ }	Yao, F., M\"{u}ller, H.-G., and Wang, J.-L. (2005). Functional linear regression analysis for longitudinal data. \textit{The Annals of Statistics}, \textbf{33(6)}, 2873--2903. 

	\bibitem{ } Zhou, L., Huang, J. Z., and Carroll, R. J. (2008). Joint modelling of paired sparse functional data using principal components. \textit{Biometrika}, \textbf{95(3)}, 601--619.
	
	\bibitem{ } Zou, H. (2006). The adaptive lasso and its oracle properties. \textit{Journal of the American Statistical Association}, \textbf{101(476)}, 1418--1429.


\end{thebibliography}

\end{document}